\documentstyle[emulateapj,epsf]{article}

\def\ltsima{$\; \buildrel < \over \sim \;$}
\def\simlt{\lower.5ex\hbox{\ltsima}}
\def\gtsima{$\; \buildrel > \over \sim \;$}
\def\simgt{\lower.5ex\hbox{\gtsima}}
 
\received{}
\accepted{}
\journalid{}{}
\articleid{}{}
\slugcomment{to appear in ApJL}
\lefthead{}
\righthead{}

\begin{document}
   
\title{Interaction in Abell 2256: the BeppoSAX view}
\author{Silvano Molendi\altaffilmark{1}, Sabrina De Grandi\altaffilmark{2} \&
Roberto Fusco-Femiano\altaffilmark{3} 
}
\altaffiltext{1}{Istituto di Fisica Cosmica, CNR, via Bassini 15,
I-20133 Milano, Italy}

\altaffiltext{2}{Osservatorio Astronomico di Brera, via Bianchi 46,
I-23807 Merate (LC), Italy}

\altaffiltext{3}{Istituto di Astrofisica Spaziale, CNR,
   via del Fosso del Cavaliere,
   I-00133 Roma, Italy}

\begin{abstract}
We present results from a spatially resolved spectral 
analysis of the merging cluster Abell 2256. 
The long integration time (135 ks) and the good spatial
resolution of the MECS onboard BeppoSAX allow us to derive a 
new and substantially improved measurement of the temperature 
structure.
We find that, within a central region of the cluster,
where the effects of the merger are visible in the ROSAT surface 
brightness image, the azimuthally averaged projected temperature
is remarkably flat; outside this region the projected temperature 
rapidly declines.
The BeppoSAX data also shows clear evidence of an azimuthal 
temperature gradient in the 4$^\prime$-8$^\prime$ (0.4-0.8 Mpc) 
radial bin, oriented in the same direction as the merger itself.
Our metal abundance profile shows, for the first time, firm evidence
of an abundance gradient in a rich merging cluster. Intriguingly 
the abundance map shows, in the SE sector, i.e. the one furthest
away from the merger, a sharp factor of two drop in abundance
at a radius comparable to the core radius of the cluster.
A possible interpretation is that, prior to the merger event, 
a cooling flow had already developed in the core of 
the infalling subgroup, as suggested by Fabian and Daines (1991).
The interaction between the substructures would have disrupted the 
cooling flow thereby re-heating and re-mixing the gas. 
As the merger in A2256 is still in a relatively early 
stage, the gas located on the side opposite to the merger event 
would still retain the low abundances associated with the 
intra-cluster medium (ICM) prior to the cooling flow disruption.

\end{abstract}


\section {Introduction}

Abell 2256 (hereafter A2256) 
is a rich, nearby  (z$=$ 0.057, Bothun \& Schombert 1990), cluster of 
galaxies.
Studies in the optical band have shown that the velocity dispersion is 
quite large ($\sim $1400 km/s; Fabricant, Kent \& Kurtz 1989, 
Bothun \& Schombert 1990). 
An early ROSAT PSPC image of A2256 (Briel et al. 1991) provided clear 
evidence of substructure, showing two emission peaks separated by about 
3.5 arcminutes. One of the two peaks is coincident with the cD galaxy 
while the distorted morphology of the other indicates that it is merging 
with the main cluster. A reanalysis, by Briel et 
al. (1991), of the velocity distribution of the galaxies measured by  
Fabricant, Kent \& Kurtz (1989) shows that it can be separated into two
distinct distributions coincident with the two X-ray peaks.  
Fabian \& Daines (1991), from the ROSAT PSPC surface brightness distribution, 
have estimated cooling times of 2$\times 10^{10}$ years and 5$\times 10^{9}$ 
years at the center of the main cluster and of the infalling subcluster 
respectively. The above authors imply that, prior to the merger event, 
a cooling flow had already developed in the core of the infalling subgroup
and that the merger may have interrupted the cooling flow and stirred up the 
gas within it.

Various attempts have been made to measure the temperature structure
of A2256. Briel \& Henry (1994), using ROSAT PSPC data, find evidence 
that the infalling group has a lower temperature than the main peak.
They also find evidence 
of two hot spots opposite each other and perpendicular to 
the presumed infall direction of the subgroup, however this result 
was not  confirmed by Markevitch \& Vikhlinin (1997) who reanalyzed 
the same data. 
Markevitch (1996, hereafter M96), from ASCA data, finds evidence of a 
smoothly declining radial temperature profile, going from $\sim$8.7 keV near 
the core to $\sim$4 keV in the outskirts.
His temperature map shows that the subgroup has a smaller 
temperature than the main peak.  
Irwin, Bregman \& Evrard (1999), from ROSAT PSPC hardness ratios, 
find a  radial profile consistent with a constant
temperature out to 15$^{\prime}$ from the cluster core. 
Their hardness ratio two dimensional map is in general agreement with 
the one of Briel \& Henry (1994).
White (1999, hereafter W99), from a  reanalysis of the ASCA data,
finds a radial temperature profile consistent with being 
constant out to 18$^{\prime}$ from the cluster core. 

In this Letter we report BeppoSAX observations of A2256.
We use our data to perform an independent measurement of the 
temperature profile and two-dimensional map of A2256.
We also present the  abundance profile and the first abundance map of 
A2256. 
The outline of the Letter is as follows.
In section 2 we give some information on the BeppoSAX observation
of A2256 and on the data preparation.
In section 3  we present spatially resolved measurements
of the temperature and metal abundance.
In section 4 we discuss our results and compare them to previous
findings.
Throughout this Letter we assume H$_{o}$=50 km s$^{-1}$Mpc$^{-1}$
and q$_{o}$=0.5.
 
\section {Observation and Data Preparation}
The cluster A2256  was observed by the BeppoSAX 
satellite (Boella et al. 1997a) at two different epochs;
between the 11$^{th}$ and the 12$^{th}$ of February 1998
and between the 25$^{th}$ and the 26$^{st}$ of February 1999.
We will discuss here data from the
MECS instrument onboard BeppoSAX; a joint analysis of the 
MECS and PDS spectra of A2256 is presented in Fusco-Femiano
et al. (2000).
The MECS (Boella et al. 1997b) is presently composed 
of two units working in the 1--10 keV 
energy range. At 6~keV, the energy resolution is  $\sim$8\%  and the 
angular resolution 
is $\sim$0.7$^{\prime}$ (FWHM). 
Standard reduction procedures and screening criteria have been
adopted to produce linearized and equalized event files.
Data preparation and linearization was performed using 
the {\sc Saxdas} package under {\sc Ftools} environment.
The total effective exposure time for the two observation was
1.3$\times$10$^5$ s. 
All spectral fits have been performed using XSPEC Ver. 10.00.
Quoted confidence 
intervals are 68\% for 1 interesting parameter (i.e. $\Delta \chi^2 =1$), 
unless otherwise stated.

\section{Spatially Resolved Spectral Analysis} 
Spectral distortions introduced by the energy dependent PSF
must be accounted for when performing spatially resolved
spectroscopy of galaxy clusters.
As for the analysis of other BeppoSAX observations of clusters
(e.g. A2319, Molendi et al. 1999), we
have taken them into account using the {\sc Effarea} program
publicly available within the latest {\sc Saxdas} release.
We remark that we fit spectra individually. This is not what 
is typically done when performing spatially resolved spectroscopy 
of clusters with ASCA data.  Here spectra accumulated from 
different regions are typically analyzed simultaneously, 
the reason being that the correction to be applied to a 
given region depends on the temperature of all the others.
The lack of a strong dependence of the MECS PSF on energy allows
us to avoid such complications.

\subsection{Radial Profiles}
For each of the two observations
we have accumulated spectra from 6 annular regions 
centered on the main X-ray emission peak of A2256, 
with inner and outer radii of 0$^{\prime}$-2$^{\prime}$, 
2$^{\prime}$-4$^{\prime}$, 4$^{\prime}$-6$^{\prime}$, 
6$^{\prime}$-8$^{\prime}$, 8$^{\prime}$-12$^{\prime}$ and  
12$^{\prime}$-16$^{\prime}$.
We have also accumulated a global spectrum from a circle
with radius 16$^{\prime}$.
The background subtraction has been performed
using spectra extracted from blank sky event files in the same region
of the detector as the source.
A correction for the absorption caused by the strongback supporting
the detector window has been applied for  the  
8$^{\prime}$-12$^{\prime}$ annulus, where the annular part of 
the strongback is contained. For the 6$^{\prime}$-8$^{\prime}$ and 
12$^{\prime}$-16$^{\prime}$ annuli, where the strongback
covers only a small fraction of the available area, we have chosen
to exclude the regions shadowed by the strongback.
For the 5 innermost annuli the energy range considered 
for spectral fitting was 2-10 keV; 
for the outermost annulus,
the fit was restricted to the  2-8 keV energy range 
to limit spectral distortions which could be caused by an incorrect 
background subtraction (see De Grandi \& Molendi 1999a for details).
Source and background spectra accumulated for each of the 
two observations have then been summed together.

We have fitted each spectrum with a MEKAL model absorbed 
by the Galactic line of sight equivalent hydrogen 
column density, $N_H$, of 4.1$\times 10^{20}$ cm$^{-2}$. 
The temperature and abundance we derive from the global
spectrum are respectively 7.5$\pm 0.1$ keV and 0.25$\pm 0.02$, 
solar units.
In figure 1  we show the temperature and abundance 
profiles obtained from our six annular regions.  
A constant does not provide a good fit to the temperature or the
abundance profile (see table 1).
   
As in Molendi et al. (1999), we have used the Fe K$_{\alpha}$ line 
as an independent estimator of the ICM temperature. 
Considering the limited number of counts available in the line, we 
have performed the analysis on 2 annuli with bounding radii,
0$^{\prime}$-8$^{\prime}$ and 8$^{\prime}$-12$^{\prime}$, the very small 
Fe abundance measured in the  12$^{\prime}$-16$^{\prime}$ annulus
prevents us from deriving a reliable line centroid for this region.
We have fitted each spectrum with a bremsstrahlung model plus a 
line, both at a redshift of z=0.057 (ZBREMSS and ZGAUSS models 
in XSPEC), absorbed by the galactic  $N_{H}$.
A systematic negative shift of 40 eV has been included in the
centroid energy to account for a slight misscalibration of 
the energy pulseheight-channel relationship near the Fe line.  
To convert the energy centroid  into a temperature 
we have derived an energy centroid vs. temperature relationship. 
This has been done by simulating thermal spectra, using the MEKAL 
model and the MECS response matrix, and fitting them with the same 
model, which has been used to fit the real data.
We derive a temperature of 8.0$^{+0.9}_{-1.0}$ keV for the inner
radial bin and of 3.2 $^{+2.8}_{-1.7}$ keV for the outer one.
Thus, our two independent measurements of the temperature 
profile are in good agreement with each other. 


\subsection{Maps}

As shown in figure 2, we have divided the MECS image of A2256 into 
4 sectors: NW, SW, SE and NE, each sector has been divided into 4 
annuli with bounding radii,
2$^{\prime}$-4$^{\prime}$, 4$^{\prime}$-8$^{\prime}$, 
8$^{\prime}$-12$^{\prime}$ and 12$^{\prime}$-16$^{\prime}$. 
The background subtraction has been performed
using spectra extracted from blank sky event files in the same region
of the detector as the source.
Correction or exclusion of the regions shadowed by the 
strongback supporting the detector window have been 
performed as in the previous subsection.
The energy ranges and the spectral models adopted for fitting are the same 
used for the azimuthally averaged spectra.

In figures 3 and 4 we show respectively  the temperature and abundance 
profiles obtained from the spectral fits for each of the 4 sectors.
In table 1 we report the best fitting constant temperatures and 
abundances for the profiles shown in figures 3 and 4.
Note that in all the profiles we have included the  
measure obtained for the central circular region with 
radius 2$^{\prime}$. 
All sectors, except for the SW sector, show a statistically significant 
temperature decrease with increasing radius.
In the NW sector
the temperature decreases continuously as the distance from the cluster
center increases. In the SE and NE sectors the temperature first increases, 
reaching a maximum in either the second (NE sector) or third (SE sector) 
annulus, and then decreases. 
Interestingly, a fit to the temperatures of the  4 sectors in the third 
annulus (bounding radii 4$^\prime$-8$^\prime$) with a constant, yields
$\chi^2=19.2$ for 3 d.o.f., with an associated probability for 
the temperature to be constant of $2.5\times 10^{-3}$, indicating that 
an azimuthal temperature gradient is present near the core of the cluster.
More specifically the NW sector of the cluster is  
the coldest, 6$\pm 0.3$ keV, and the SE sector the hottest, 8.4$\pm 0.5$ keV.
%
The SE sector is the only one to show clear evidence of an abundance decline 
with increasing radius, all other sectors have abundance profiles which
are consistent with being flat.

\begin{table*}
\caption{Best fit constant values for the temperature and
the abundance for the azimuthally averaged radial
profiles (All) and for the profiles of the 4 sectors (NW, SW, SE and NE)}
\centerline{
\begin{tabular}{lccccccc}
\hline
Sector &    kT       & $\chi^2$ & Prob.$^a$         & Abundance   & $\chi^2$ & Prob.$^a$         &d.o.f. \cr
       &  (keV)      &          &                   &(Solar Units)&          &                   &    \cr
\hline
\hline
All    & 6.9$\pm$0.1 &   36.4   &$7.9\times 10^{-7}$& 0.25$\pm$0.02&   13.9  &$1.6\times 10^{-2}$& 5  \cr  
\hline
NW     & 6.6$\pm$0.2 &   11.3   &$2.4\times 10^{-2}$& 0.30$\pm$0.03&    3.8  & 0.43              & 4  \cr
SW     & 7.0$\pm$0.2 &    7.4   & 0.12              & 0.27$\pm$0.03&    8.1  &$9.0\times 10^{-2}$& 4  \cr
SE     & 7.2$\pm$0.2 &   11.3   &$2.4\times 10^{-2}$& 0.21$\pm$0.03&   29.0  &$7.8\times 10^{-6}$& 4  \cr
NE     & 6.7$\pm$0.2 &   52.2   &$1.3\times10^{-10}$& 0.30$\pm$0.03&    2.0  & 0.74              & 4  \cr
\hline
\end{tabular}
}
\par
$^a$Probability for  the observed distribution to be drawn from a constant parent distribution.
\end{table*}

\section{Discussion}

Previous measurements of the temperature structure of A2256
have been performed by Briel \& Henry (1994) and Irwin et al. (1999),
using ROSAT data, by  M96 and by W99 using ASCA data. 
We have performed a detailed comparison of our radial temperature 
profile with the ones based on the ASCA satellite (M96 and W99) 
which covers an energy  range similar to ours.
In figure 1 we have overlaid the temperature profile obtained by  
M96 and by W99 on our profile.
The higher quality of the BeppoSAX measurement, due in part
to the much longer exposure time and in part to the better
angular resolution of our instrument, is quite evident.
The innermost bin in the M96 profile, 0$^\prime$-6$^\prime$ has
a temperature that is inconsistent with the temperature 
we measure from our three innermost bins spanning the same radial range.
It must be noted that, while our profile is azimuthally averaged over 
all angles, the M96 measurement has been obtained excluding the region 
presumably contaminated by the softer emission of the infalling group.
To obtain a direct comparison between our measurement and the one reported
in M96, we have derived the temperature from a circular region with 
radius 6$^\prime$ excluding the NW sector containing the infalling 
group. Our measurement, 7.5$\pm$0.2 keV, although somewhat higher than the 
one obtained by averaging over all directions, is still incompatible at more
than the 3$\sigma$ level with the one reported by M96.
The second radial bin reported in M96
 (6$^\prime$-11$^\prime$) is characterized by a temperature 
apparently larger than the mean temperature for our corresponding bins 
(i.e.  6$^\prime$-8$^\prime$ and 8$^\prime$-12$^\prime$).
However this difference is only apparent, indeed if we simultaneously
fit the BeppoSAX spectra for the 6$^\prime$-8$^\prime$ and 
8$^\prime$-12$^\prime$ bins, which is equivalent to fitting data
from the 6$^\prime$-12$^\prime$ bin, we derive a temperature of 
6.6$\pm$0.3 keV, which is consistent with the one derived by M96.
The temperature for the outermost bin in the M96 profile is in  
agreement with our own measurement. 
The  W99 measurement, which comes from a
different analysis of the same ASCA observation used by M96,
is in agreement with ours for radii smaller than 6$^\prime$.
The outermost bin reported in W99 appears to have a temperature 
substantially larger than the mean temperature for our corresponding bins 
(i.e.  6$^\prime$-8$^\prime$, 8$^\prime$-12$^\prime$ and 
12$^\prime$-16$^\prime$). This difference is only apparent, if we 
simultaneously fit the BeppoSAX spectra for the 6$^\prime$-8$^\prime$, 
the 8$^\prime$-12$^\prime$ and the 12$^\prime$-16$^\prime$ bins we derive 
a temperature of 6.5$\pm$0.3,
which is consistent with the one derived by W99.
The apparent difference is related to the strong gradient in the surface 
brightness profile when going from 6$^\prime$ to 16$^\prime$, which causes 
the emission from the entire region to be dominated 
by the contribution of the innermost annuli.
In summary: for radii larger than 6$^\prime$, our profile is in agreement 
with the M96 and W99 profile, while for  radii smaller than 6$^\prime$,
our profile is in agreement with the  W99 profile and
in disagreement with the M96 profile.

The most striking feature of our radial temperature profile is the 
presence of  a relatively localized gradient. The temperature is 
flat out to 8$^\prime$ and decreases by almost a factor
two within the following 8$^\prime$.
The radius at which the temperature starts to decline
$\sim 8^\prime$ (0.8 Mpc) is comparable to the radius at which the 
X-ray isophotes are no longer disturbed by the interaction of the two
subclusters, which is clearly seen at smaller scales in the ROSAT PSPC 
image (e.g. figure 2 of Briel et al. 1991).
Thus the presence of a hot almost isothermal region in the core is most 
likely related  to the on-going merger between the main cluster and the 
group. 
The BeppoSAX temperature map shows clear evidence of an azimuthal gradient 
in the 8$^\prime$-12$^\prime$ radial bin. The NW sector is found to be the 
coldest while the SE sector appears to be the hottest, thus the 
gradient appears to be oriented in the same direction as the merger itself. 
Interstingly, in a previous work (De Grandi \& Molendi 1999a), the merging 
cluster A3266 was found to have a similar temperature structure.
No evidence of the two hot spots reported by Briel \& Henry (1994)
is found in our map. 
 
The metal abundance in A2256 appears to decrease with increasing radius
(see figure 1). 
This is the first firm case, to our knowledge, of an abundance 
gradient in a rich non cooling flow cluster.
Evidence of an abundance gradient has been found in the poor cluster 
MKW4 (Finoguenov et al. 1999), while marginal evidence has been
found in A399 (Fujita et al. 1996) and A1060 (Finoguenov et al.
1999). 
In A2256 the abundance averaged over a central region of 0.2 Mpc 
radius is 
$\sim$ 0.3, solar units, a  value which, although higher than the average 
abundance for non cooling flow cluster, $\sim$ 0.20 (Allen \& Fabian 1998),
is smaller than those commonly observed in the core of cooling flow cluster 
(see for example Finoguenov et al. 1999, for an
analysis of abundance profiles from ASCA data, and our own BeppoSAX 
results on Abell 2029, Molendi \& De Grandi 1999, and PKS 0745-191,
De Grandi \& Molendi 1999b).
Furthermore, the abundance map (see figure 4) shows that the SE sector, 
i.e. the one furthest away from the on-going merger, presents a highly 
significant abundance decline  (probability $= 7.6\times 10^{-6}$)
localized at a radius comparable to the core radius of the cluster.
A possible interpretation is that, prior to the merger event, 
a cooling flow had already developed in the core
of the infalling subgroup, as suggested by Fabian \& Daines (1991). 
The above authors, from the gas densities
at the center of the main cluster and of the infalling subcluster
compute cooling times of 2$\times 10^{10}$ years and 
5$\times 10^{9}$ years respectively, implying that the 
infalling subcluster must have had a cooling flow. 
The interaction between the substructures would have disrupted the 
cooling flow thereby re-heating and re-mixing the gas. 
As the merger in A2256 is still in a relatively early 
stage, the gas located on the side opposite to the merger event 
may still retain the low abundances associated to the 
ICM prior to the cooling flow disruption.
It seems unlikely that a contribution to the metallicity 
enhancement has come from the main cluster as its core 
density implies a cooling time that is larger than the age of the
Universe. 
Finally we speculate that other rich merging clusters,
similar to A2256, may present metallicity gradients produced 
by disrupted cooling flows. BeppoSAX and future XMM
observations of merging clusters will certainly contribute in clarifying 
this issue.


\acknowledgments
We acknowledge support from the BeppoSAX Science Data Center.
We thank the referee for useful comments and D. Lazzati for 
help in producing the contour plot.  


\clearpage


{\begin{figure}
\epsfxsize=\textwidth
\epsffile{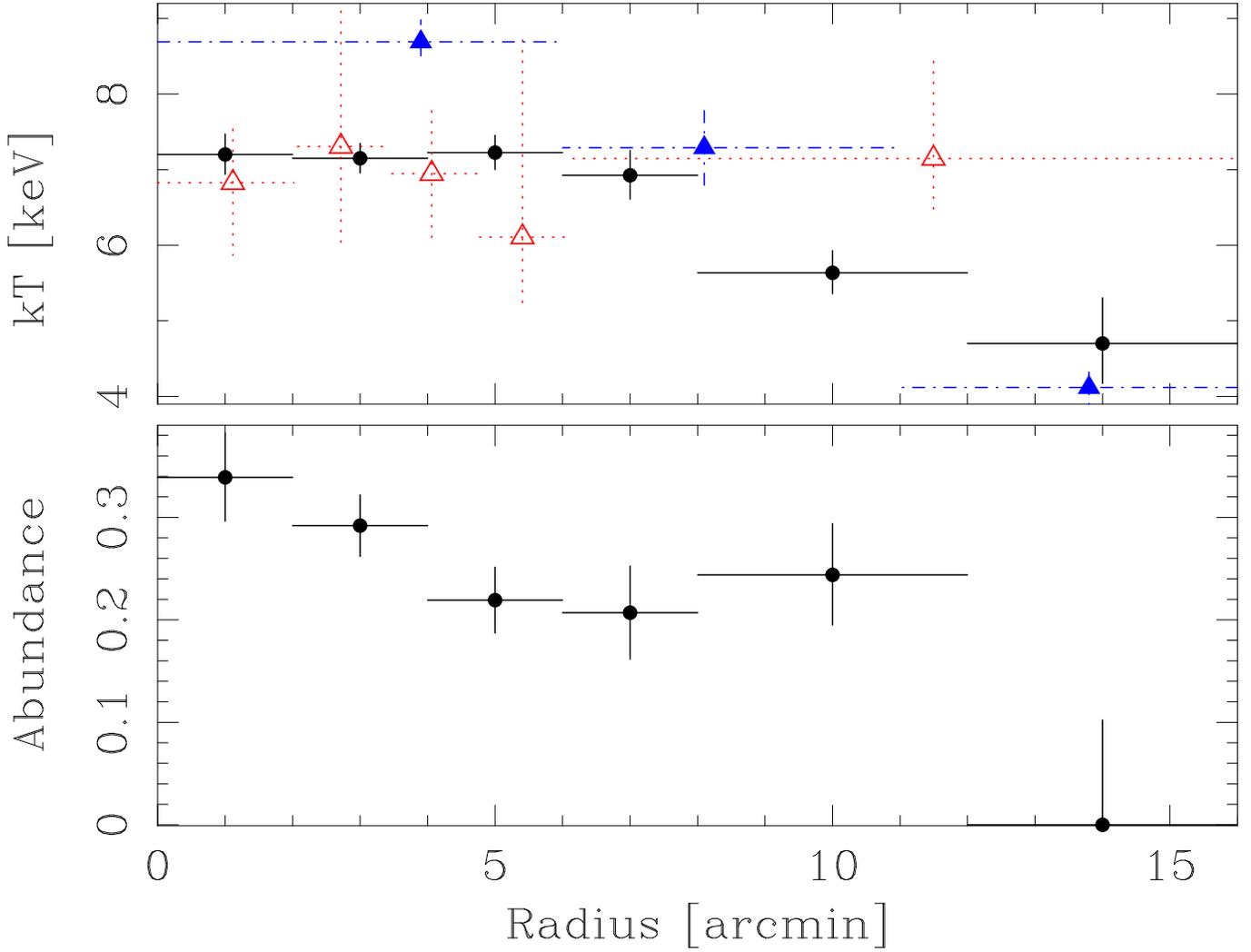}
\vskip -11cm
\figcaption
{{\bf Top Panel}: Projected radial temperature profile for A2256.
Filled circles indicate temperatures obtained from BeppoSAX MECS data.
Filled and open triangles represent the temperature profile derived 
from ASCA data, respectively by M96 and W99.
All the uncertainties on the
temperature measurements are at the 68$\%$ confidence level (we have
converted the 90$\%$ confidence errors reported in figure 4 of
Markevitch et al. 1998 into 68$\%$ confidence errors by dividing
them by 1.65, for further details see Markevitch \& Vikhlinin 1997).  
{\bf Bottom Panel}: projected radial abundance profile from BeppoSAX MECS
data. 
}
\end{figure}}
\clearpage

\begin{figure}
\vskip -6cm
\plotone{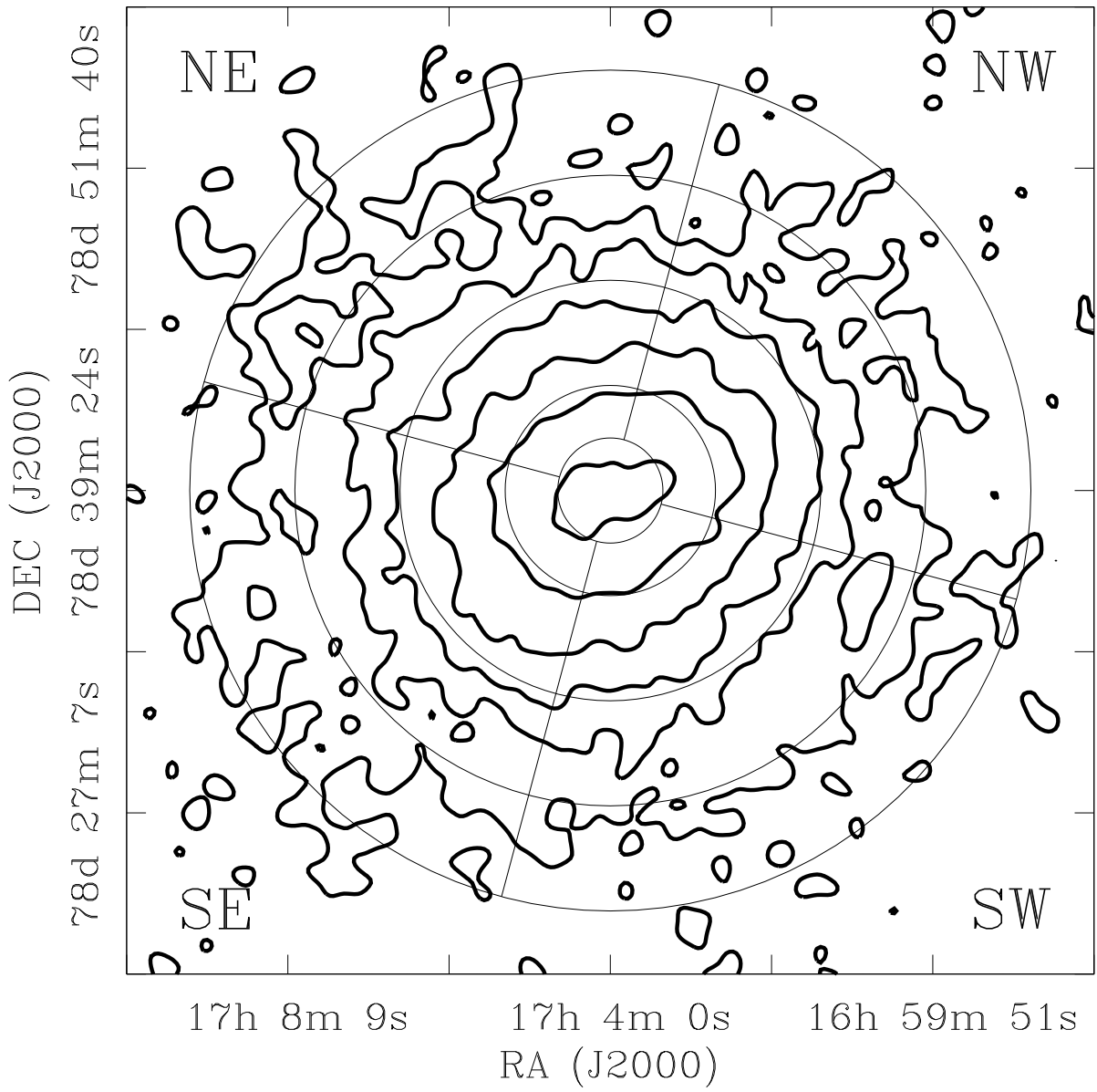}
\vskip 2cm
\caption
{BeppoSAX MECS image of A2256. Contour levels are indicated
by the thick lines. The thin lines show how the cluster
has been divided to obtain temperature and abundance maps.
}
\end{figure}
\clearpage

\begin{figure}
\vskip -1cm
\plotone{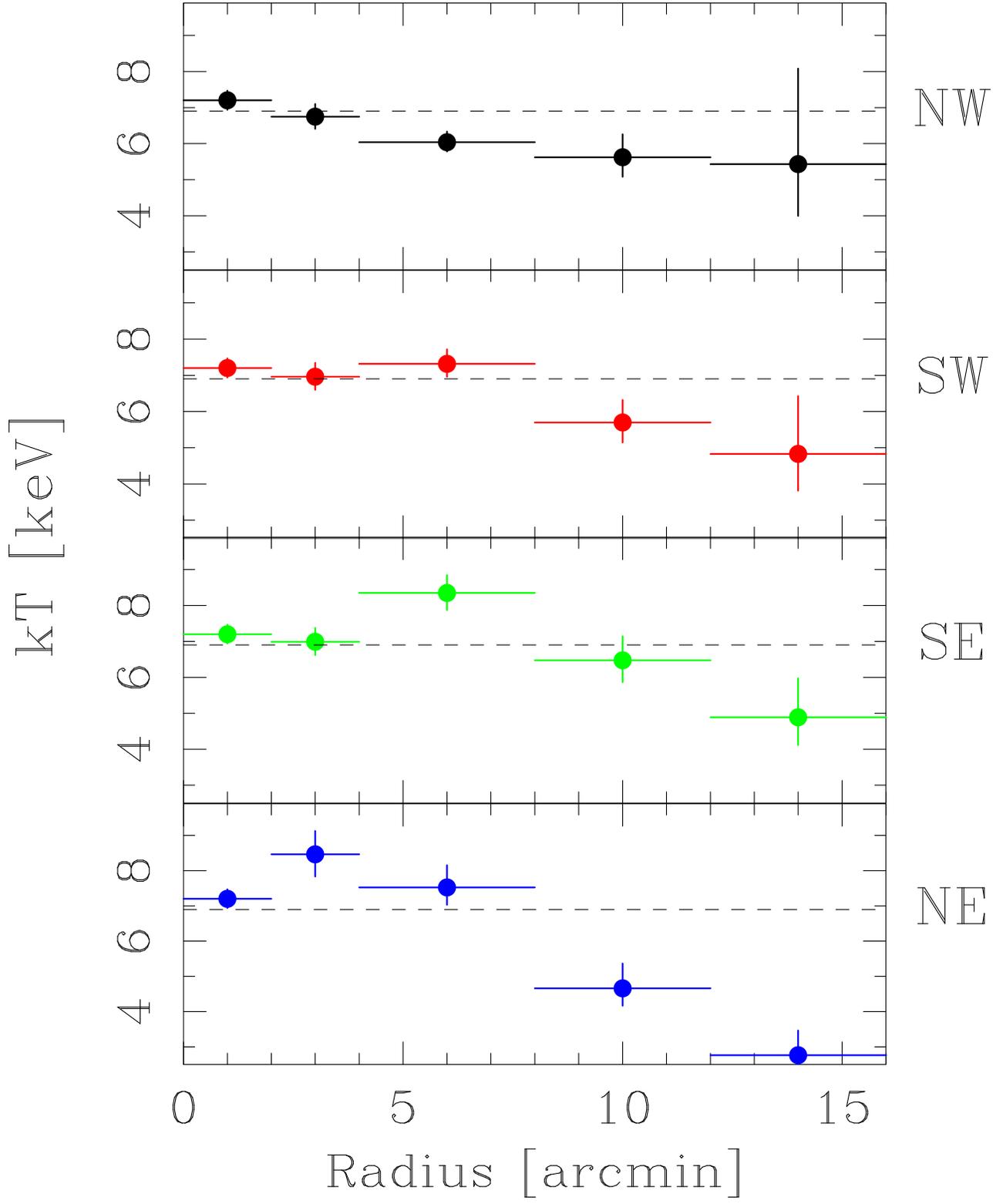}
\vskip -1cm
\caption
{Radial temperature profiles for the NW sector (first panel), the SW sector
(second panel), the SE sector (third panel) and the NE sector (forth
panel).
The temperature for the leftmost bin is derived from the entire circle,
rather than from each sector. The dashed lines indicate the average 
temperature derived from the MECS temperature profile reported in 
figure 1.
}
\end{figure}
\clearpage

\begin{figure}
\vskip -1cm
\plotone{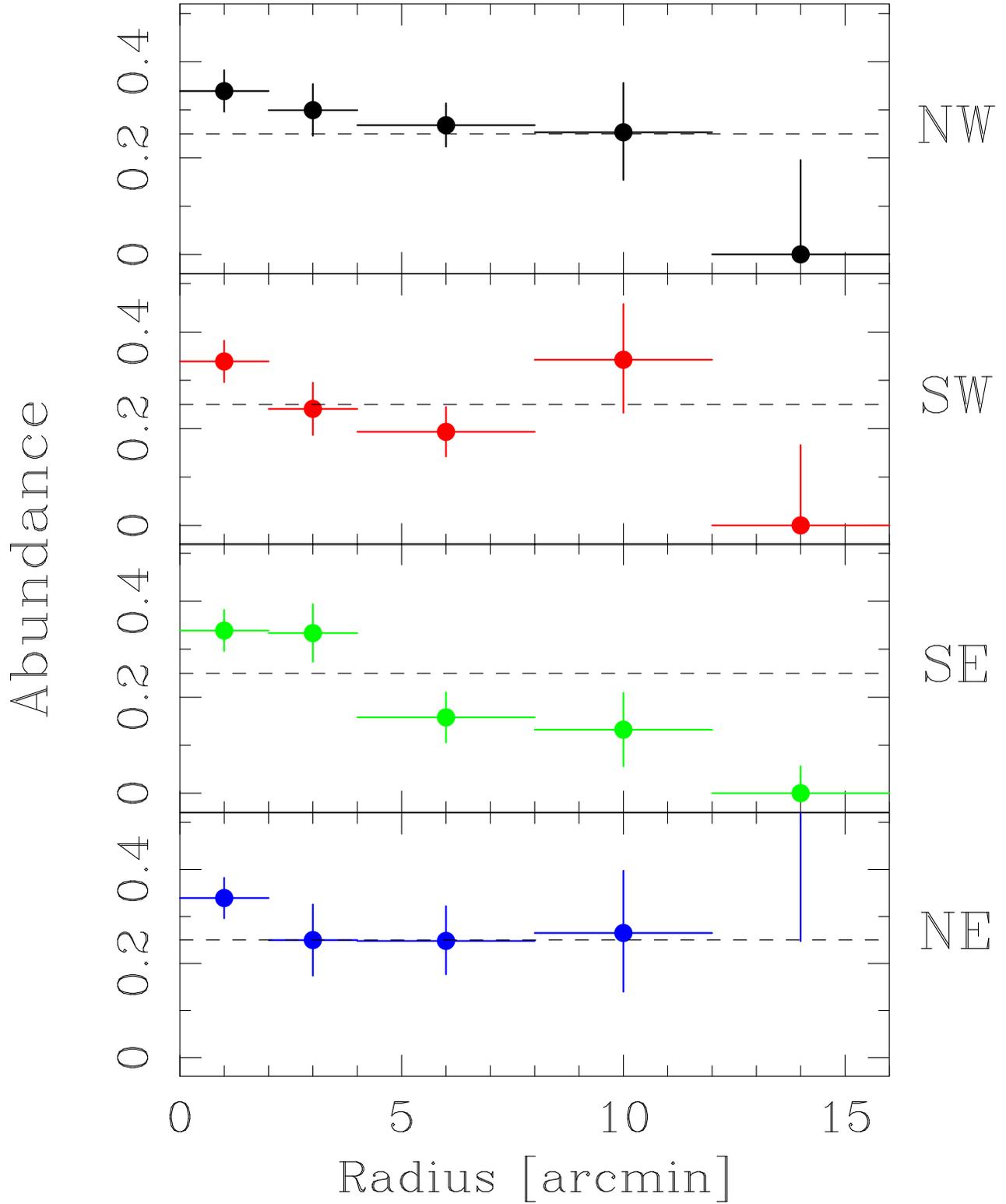}
\vskip -1cm
\caption
{Radial abundance profiles for the NW sector (first panel), the SW sector
(second panel), the SE sector (third panel) and the NE sector (forth
panel).
The abundance for the leftmost bin is derived from the entire circle,
rather than from each sector. 
The dashed lines indicate the average 
abundance derived from the profile reported in 
figure 1.
The abundance measure for the 8$^\prime$-12$^\prime$ bin in the NE
sector is 0.89$^{+1.40}_{-0.64}$. 
}
\end{figure}


\end{document}